\documentclass[sigconf]{acmart}
\usepackage{amsmath,amsfonts}
\usepackage{algorithmic}
\usepackage{algorithm}
\usepackage{array}
\usepackage[caption=false,font=normalsize,labelfont=sf,textfont=sf]{subfig}
\usepackage{textcomp}
\usepackage{stfloats}
\usepackage{url}
\usepackage{verbatim}
\usepackage{graphicx}
\usepackage{graphics}

\usepackage{multirow}
\usepackage{rotating}
\begin{document}
	
	\title{MD2PR: A Multi-level Distillation based Dense Passage Retrieval Model}
	
	\author{Haifeng Li}
%	\authornote{Both authors contributed equally to this research.}
	\email{mydlhf@cufe.edu.cn}
	\orcid{1234-5678-9012}
	\author{Mo Hai}
	\authornote{Corresponding Author.}
%	\authornotemark[1]
	\email{haimo@cufe.edu.cn}
	\author{Dong Tang}
	%	\authornotemark[1]
	\email{tangdong@cufe.edu.cn}
	\affiliation{%
		\institution{School of Information, Central University of Finance and Economics}
		\streetaddress{39 South College Road, Haidian District}
		\city{Beijing}
%		\state{Beijing}
		\country{China}
		\postcode{100081}
	}
	
%	\author{Lars Th{\o}rv{\"a}ld}
%	\affiliation{%
%		\institution{The Th{\o}rv{\"a}ld Group}
%		\streetaddress{1 Th{\o}rv{\"a}ld Circle}
%		\city{Hekla}
%		\country{Iceland}}
%	\email{larst@affiliation.org}
%	
%	\author{Valerie B\'eranger}
%	\affiliation{%
%		\institution{Inria Paris-Rocquencourt}
%		\city{Rocquencourt}
%		\country{France}
%	}
%	
%	\author{Aparna Patel}
%	\affiliation{%
%		\institution{Rajiv Gandhi University}
%		\streetaddress{Rono-Hills}
%		\city{Doimukh}
%		\state{Arunachal Pradesh}
%		\country{India}}
%	
%	\author{Huifen Chan}
%	\affiliation{%
%		\institution{Tsinghua University}
%		\streetaddress{30 Shuangqing Rd}
%		\city{Haidian Qu}
%		\state{Beijing Shi}
%		\country{China}}
%	
%	\author{Charles Palmer}
%	\affiliation{%
%		\institution{Palmer Research Laboratories}
%		\streetaddress{8600 Datapoint Drive}
%		\city{San Antonio}
%		\state{Texas}
%		\country{USA}
%		\postcode{78229}}
%	\email{cpalmer@prl.com}
%	
%	\author{John Smith}
%	\affiliation{%
%		\institution{The Th{\o}rv{\"a}ld Group}
%		\streetaddress{1 Th{\o}rv{\"a}ld Circle}
%		\city{Hekla}
%		\country{Iceland}}
%	\email{jsmith@affiliation.org}
%	
%	\author{Julius P. Kumquat}
%	\affiliation{%
%		\institution{The Kumquat Consortium}
%		\city{New York}
%		\country{USA}}
%	\email{jpkumquat@consortium.net}
	
	\renewcommand{\shortauthors}{Haifeng et al.}
	
	\begin{abstract}
	Ranker and retriever are two important components in  dense passage retrieval. The retriever typically adopts a dual-encoder model, where queries and documents are separately input into two pre-trained models, and the vectors generated by the models are used for similarity calculation. The ranker often uses a cross-encoder model, where the concatenated query-document pairs are input into a pre-trained model to obtain word similarities. However, the dual-encoder model lacks interaction between queries and documents due to its independent encoding, while the cross-encoder model requires substantial computational cost for attention calculation, making it difficult to obtain real-time retrieval results. In this paper, we propose a dense retrieval model called MD2PR based on multi-level distillation. In this model, we distill the knowledge learned from the cross-encoder to the dual-encoder at both the sentence level and word level. Sentence-level distillation enhances the dual-encoder on capturing the themes and emotions of sentences. Word-level distillation improves the dual-encoder in analysis of word semantics and relationships. As a result, the dual-encoder can be used independently for subsequent encoding and retrieval, avoiding the significant computational cost associated with the participation of the cross-encoder. Furthermore, we propose a simple dynamic filtering method, which updates the threshold during multiple training iterations to ensure the effective identification of false negatives and thus obtains a more comprehensive semantic representation space. The experimental results over two standard datasets show our MD2PR outperforms 11 baseline models in terms of $MRR$ and $Recall$ metrics.
	\end{abstract}
	
%	\begin{CCSXML}
%		<ccs2012>
%		<concept>
%		<concept_id>00000000.0000000.0000000</concept_id>
%		<concept_desc>Do Not Use This Code, Generate the Correct Terms for Your Paper</concept_desc>
%		<concept_significance>500</concept_significance>
%		</concept>
%		<concept>
%		<concept_id>00000000.00000000.00000000</concept_id>
%		<concept_desc>Do Not Use This Code, Generate the Correct Terms for Your Paper</concept_desc>
%		<concept_significance>300</concept_significance>
%		</concept>
%		<concept>
%		<concept_id>00000000.00000000.00000000</concept_id>
%		<concept_desc>Do Not Use This Code, Generate the Correct Terms for Your Paper</concept_desc>
%		<concept_significance>100</concept_significance>
%		</concept>
%		<concept>
%		<concept_id>00000000.00000000.00000000</concept_id>
%		<concept_desc>Do Not Use This Code, Generate the Correct Terms for Your Paper</concept_desc>
%		<concept_significance>100</concept_significance>
%		</concept>
%		</ccs2012>
%	\end{CCSXML}
	
%	\ccsdesc[500]{Do Not Use This Code~Generate the Correct Terms for Your Paper}
%	\ccsdesc[300]{Do Not Use This Code~Generate the Correct Terms for Your Paper}
%	\ccsdesc{Do Not Use This Code~Generate the Correct Terms for Your Paper}
%	\ccsdesc[100]{Do Not Use This Code~Generate the Correct Terms for Your Paper}
	
%	\begin{teaserfigure}
%		\includegraphics[width=\textwidth]{sampleteaser}
%		\caption{Seattle Mariners at Spring Training, 2010.}
%		\Description{Enjoying the baseball game from the third-base
%			seats. Ichiro Suzuki preparing to bat.}
%		\label{fig:teaser}
%	\end{teaserfigure}
	
%	\received{20 February 2007}
%	\received[revised]{12 March 2009}
%	\received[accepted]{5 June 2009}
	
	\maketitle

\section{Introduction}\label{sec:intro}

In information retrieval, the main task is to match the query and the documents. Generally, a retriever is used to retrieve a set of passages from a massive collection of documents that have relatively high relevance to a given query; these passages are then further ranked by a ranker based on their relevance scores, to obtain the most relevant passages to the query. Traditional methods employ bag-of-words models such as TF-IDF\cite{ramos2003using} and  BM25\cite{robertson2009probabilistic} to represent the similarity between the query and the documents. With the development of deep learning, word embedding has been widely used as a new representation. Both bag-of-words model and word embedding are sparse passage retrieval that rely on one-hot encoding or independent vector based on word frequencies. Sparse retrieval models have limitations such as vocabulary mismatch, semantic mismatch, and lack of contextual interaction, which make it challenged to meet the increasingly complex and diverse business needs and application scenarios. 

In recent years, deep models like Transformer\cite{vaswani2017attention} and Bert\cite{devlin2018bert} have been applied to information retrieval. They represent words and sentences as dense vectors, capturing contextual information and semantic relevance between query and documents, known as the dense passage retrieval. The advantage lies in scenarios where there is minimal overlap between the query and the documents. However, it requires labeled data to optimize model and thus performs worse than sparse retrieval model in zero-shot conditions.

The main dense retrieval models include the dual-encoder and cross-encoder models. In a dual-encoder model, two Transformer-liked models, such as Bert, are separately trained for the query and the documents, and the semantic representation of the entire sentence is obtained using a CLS token, which is used to calculate the relevance. The cross-encoder model, on the other hand, trains the query and documents together by concatenating them, enabling interaction between words in the query and passage. This model can capture more contextual information and generally performs better than the dual-encoder model. 

However, these dense retrieval methods still have 3 main problems: (1) The dual-encoder model encodes the query and passage separately, resulting in a lack of interaction between them, which can distort the similarity calculation and cause a loss of contextual information, leading to sub-optimal performance. (2) The cross-encoder model employs an attention mechanism for full interaction, but this comes at a high computational cost, which is proportional to the square of the text length\cite{kitaev2020reformer}. While it effectively improves model performance, it also significantly decrease both training and inference computational efficiency. Experimental results from COIL\cite{gao2021coil} show that using BM25 retrieval has a latency of 36 milliseconds, whereas using the dual-encoder model DPR\cite{karpukhin2020dense} increases the latency to 293 milliseconds, an 8-fold increase. The use of the cross-encoder model ColBert\cite{khattab2020colbert} further increases the latency to 458 milliseconds, nearly 13 times of the BM25. (3) Negative samples are typically randomly sampled from the current batch of data or the entire dataset, lacking true annotations. These samples may only be highly similar pseudo-negative ones to the query. Simply pushing away the query from them can impact the sentence representation and affect the overall model performance.

Current research has focused on the combined use of cross-encoder and dual-encoder models in dense passage retrieval. In this paper, we propose a multi-level distillation model called MD2PR, in which a dual-encoder model is utilized for large-scale candidate retrieval during the retrieval phase, while a cross-encoder model is employed for fine ranking of the retrieved results during the ranking phase. This method can not only improve the quality of the retrieval results but also decrease the computational cost. The main contributions are shown as follows.

\begin{enumerate}
	\item We use a knowledge distillation method, by which the knowledge learned by the ranker is transferred to the retriever at both sentence and word levels. This overcomes the lack of interaction between the query and document in the dual-encoder model, allowing the retriever to effectively find documents with higher relevance to the given query. It also reduces the computational cost introduced by the cross-encoder model.

	\item In the word-level distillation, we explore the correspondence between the attention of the ranker and the hidden states of the retriever. Through the optimization of a constructed loss function, the attention of the retriever can be exactly approximated to that of the ranker.

	\item We propose a dynamical filtering method, which uses a dynamic threshold, instead of a fixed threshold, to select the false negatives. This method reduces the bias introduced during the sampling process and ensures the consistency of the representation space.

	\item We evaluate our proposed model on two standard datasets using 11 baseline models. The experimental results show that the proposed model have an significant improvement over the classical retrieval models in various evaluation metrics.

\end{enumerate}

The organization structure of this paper is as follows: Section \ref{sec:related} reviews the relevant literature related to our research topic. Section \ref{sec:mddpr} proposes the MD2PR model and provides a detailed algorithmic description of the model. Section \ref{sec:expr} makes an experimental studies, and Section \ref{sec:conclude} concludes this paper.

\section{Related Works}\label{sec:related}

\subsection{Knowledge Distillation}

The pre-trained models has improved the performance of downstream tasks. However, it also led to the challenge of dealing with massive number of parameters during training and deployment. Knowledge distillation, proposed by Hinton\cite{hinton2015distilling}, was a transfer learning model that allows for the reduction of training and deployment complexity without significantly impacting performance. Knowledge distillation typically involves a more complex teacher model with better predictive accuracy and a simpler  student model for inference. The soft labels generated by the teacher model serve as a guidance to optimize the student model in a favorable direction, achieving the goal of knowledge transfer. The total loss in knowledge distillation consists of two parts: one is the contrastive loss between the student model and the ground truth labels, and the other is the contrastive loss between the student model and the soft labels generated by the teacher model, also known as the distillation loss. The overall loss is the weighted sum of these two parts. As a result, the larger the weight of the distillation loss, the greater the proportion of knowledge transfer induced. 

DistilBert\cite{romero2014fitnets} was an attempt to distill large models by using Bert as the teacher model. Based on duplicating the network and reducing the number of layers by half, the reduced model is trained using 3 types of losses: the cross-entropy loss between the outputs of the teacher model and the student model, the cross-entropy loss between the outputs of the student model and the ground truth labels, and the cosine similarity between the outputs of the teacher model and the student model. DistilBert was capable of reducing the model size by 40\% while retaining 97\% of the language understanding capability, and it also improved the computational speed by 60\%. Another distillation model, TinyBert\cite{jiao2019tinybert}, introduced a two-stage distillation process where the teacher model is distilled in both the pre-trained and fine-tuning stages. This ensures that TinyBERT can learn both the general domain and task-specific knowledge from Bert. TinyBERT also proposed distillation of the Transformer layers, including attention distillation and hidden layer state distillation. Attention distillation transfers language knowledge, while hidden layer state distillation uses a learnable linear transformation to convert the hidden states of the student model into a space similar to that of the teacher model.

\subsection{Dense Retrieval}

The deep language models have been increasingly applied in information retrieval, resulting in significant improvements in retrieval effectiveness. The loss function used in dense retrieval training is contrastive learning, aiming to maximize the similarity between query and positive documents while minimizing the similarity between query and negative documents.

\subsubsection{Dual-Encoder}

Dual-encoder models can be classified into single representation and multi-representation. The advantage of dual-encoder models is that they include two encoders. During prediction, the query and passage only need to be separately input to the two encoders. The CLS token is used to represent the semantics of the query and document, and their matching score is calculated using similarity methods. This approach reduces the text length of the input model and eliminates the interaction between the query and the document. The time complexity is the sum of the time complexities of the two encoders. However, since the dual-encoder model only utilizes the information from the CLS token to represent the entire sentence, the performance greatly relies on the accuracy of the CLS token. Additionally, because there is no interaction between the query and the document, the accuracy is not high enough.

DPR\cite{karpukhin2020dense} was a classic dual-encoder model, whose task was question-answering, which aimed to find the most relevant answer to a given question from a corpus of existing text. DPR utilized two Bert models as encoders to construct  representations for queries and passages. The matching scores between them are measured using dot product\cite{johnson2019billion}. DPR is also referred to as a representation-based retrieval model\cite{gao2021complement} because it represents both queries and passages and calculates their similarity using a similarity function. ANCE\cite{xiong2020approximate} also adopted a dual-encoder model, which replaced batch-level negative samples with global negative samples and updates the index asynchronously. One improvement approach for dual-encoder model is to independently train multiple encoders to represent queries or passages with multiple vectors. By summing or taking the average of these vectors, queries and passages can be more comprehensively represented without significantly increasing training costs. Among them, Poly-Encoders\cite{humeau2019poly} chose to encode queries into multiple vectors, while Me-Bert\cite{luan2021sparse} chose to encode passages into multiple vectors.

\subsubsection{Cross-Encoder}

The cross-encoder model\cite{dai2019deeper} used only one encoding model. It concatenates the query and passage and then inputs them into the model for training. The model ultimately outputs an embedded vector and calculates the relevance score. The cross-encoder model can achieve the finest-grained interaction, involving all the words between the query and the passage. It is also known as an interaction-based retrieval model\cite{gao2021complement}. The interaction allows the model to identify the matching scores between the query and the passage more accurately. However, the main challenge of the cross-encoder model lies in the high computational cost of the cross-attentions, especially during prediction, which requires concatenating each query with all the documents and inputting them into the encoding model to find the most matching passage. Thus, the time complexity is the product of the number of queries and documents. Additionally, the consideration of frequently occurring function words in the document decreases the distinction between different documents.

ColBert\cite{khattab2020colbert} introduced the late interaction to reduce computational costs while inheriting the cross-encoder model. It used prefix tokens to separately identify the query and passage. However, it only implements interaction within the query and passage until the last layer, where the interaction is achieved through a MaxSim operation. COIL\cite{gao2021coil} also adopted the dual-encoder model by referring to the interaction approach from ColBert. The difference between COIL and ColBert is that only overlapping words between the query and passage have interactions, rather than involving all words.

\subsubsection{Combination of Dual-Encoder and Cross-Encoder}

Due to the efficiency of the dual-encoder model, it is used in the retrieval phase to find a small number of passages from a massive amount of text that are most relevant to a given query. On the other hand, the accuracy of the cross-encoder model makes it suitable for ranking a small amount of text to identify the most matching passages. Therefore, using the knowledge learned by the ranker to guide the retriever is an effective approach. The Adversarial Retriever-Ranker(AR2)\cite{zhang2021adversarial} model introduces the generative adversarial network into information retrieval, drawing inspiration from the IRGAN framework\cite{wang2017irgan}. The dual-encoder retriever serves as the generator, retrieving hard negative samples to confuse the ranker. The cross-encoder ranker acts as the discriminator, learning and ranking a set of samples that include positive samples and hard negatives. Both the retriever and ranker calculate the relevance scores between the query and the passage. The retriever and ranker are fixed separately during training until the model converges through iterative training.

\subsubsection{Negative Samples}

In information retrieval, positive samples are typically manually selected, while negative samples are randomly chosen from the entire data. As a result, it is difficult to ensure that all selected negative samples are truly negative. Many researchers tried to improve the quality of negative samples. DCLR\cite{zhou2022debiased} designed an instance weighting method to penalize false negative samples and generate noise-based negative samples to ensure the uniformity of the representation space. ANCE\cite{xiong2020approximate} abandoned the conventional intra-batch negative sampling and adopted global negative sampling to reduce variance and accelerate convergence. SimANS\cite{zhou2022simans} sampled ambiguous negative samples using a new probability distribution. RocketQA\cite{qu2020rocketqa,ren2021rocketqav2} compromised between intra-batch negative sampling and global negative sampling, selecting all negative samples as candidates during parallel training. In the field of image retrieval, there have been attempts to increase the number of negative samples through data augmentation\cite{chen2020simple}. These approaches have put a lot of effort into constructing negative samples, resulting in significant computational costs.

\subsubsection{Redesigning the Pre-trained Model}

The pre-training tasks of Bert models, such as Masked Language Model or Next Sentence Prediction\cite{devlin2018bert}, are not fully suitable for information retrieval. Redesigning pre-trained models can reduce the gap between pre-training tasks and retrieval tasks. Condenser proposed a new transformer pre-trained model\cite{gao2021condenser}, which consists of 3 parts: early encoder layers, late encoder layers, and head layers. It pays more attention to the global representation of input text using the CLS token, allowing competitive results to be achieved with less data. CoCondenser\cite{gao2021unsupervised} was also a pre-trained model proposed to reduce the training difficulty of dense retrievers. In the first stage, it used Condenser for pre-training, and in the second stage, it used a gradient caching model for contrastive learning. SimLM\cite{wang2022simlm} was inspired from Condenser and compresses the semantics of paragraphs into dense vectors through pre-training.   

\section{Multi-level Distillation Dense Passage Retrieval}\label{sec:mddpr}

\begin{figure*}[h]
	\centering
	\includegraphics[height=7cm]{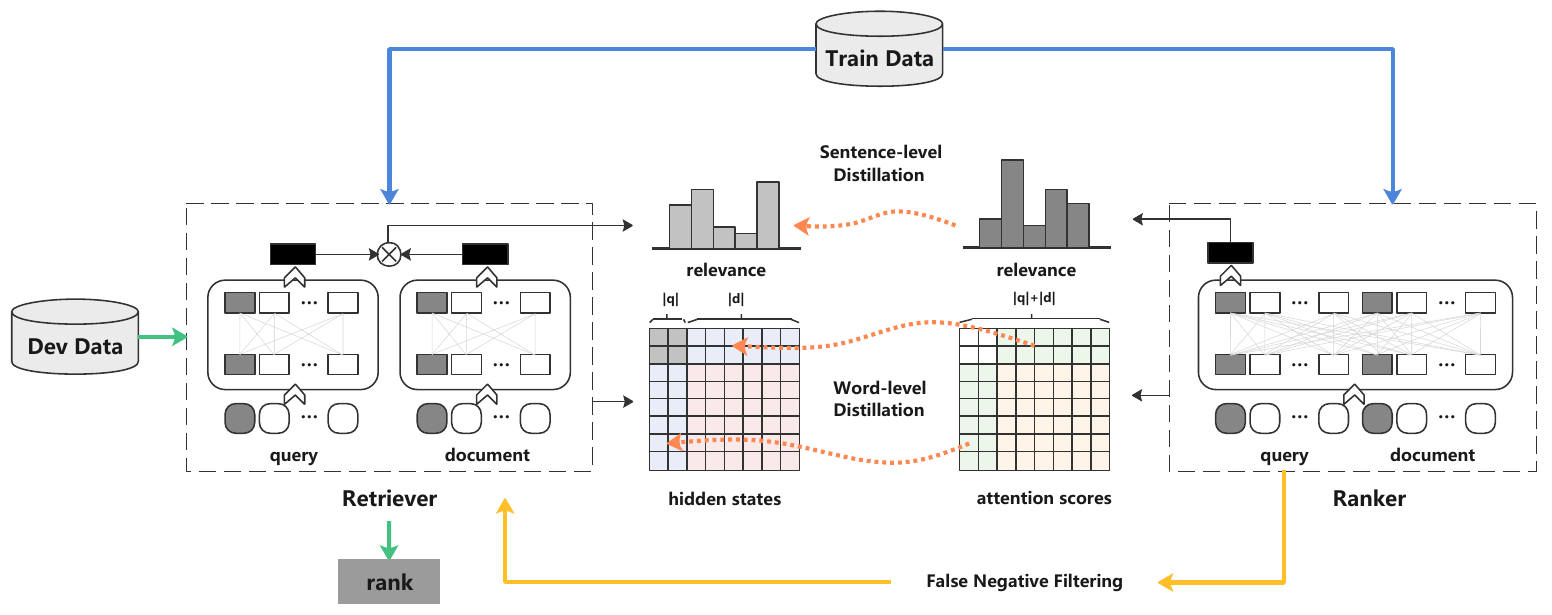}
	\caption{Illustration of MD2PR. (a) Blue line: train data is fed to the retriever and ranker, where retriever learn knowledge by distillation from ranker; (b) Yellow line: false negative filtering method is performed based on the score achieved by ranker; (c) Green line: validation data is fed to the retriever and get the rank; (d) Orange line: sentence-level and word-level distillation.}\label{fig:md2pr}
%	\Description{A woman and a girl in white dresses sit in an open car.}
\end{figure*}

In this paper, we propose a model named MD2PR(\textbf{M}ulti-level \textbf{D}istillation \textbf{D}ense \textbf{P}assage \textbf{R}etrieval), which includes sentence-level distillation and word-level distillation. Based on the knowledge distilled at these two levels, the interaction knowledge between the query and documents, learned by the cross-encoder serving as a ranker, is effectively transferred to the dual-encoder serving as a retriever. Additionally, to reduce the bias introduced by random negative sampling, a dynamic false negative filtering method is introduced. This method can improve the quality of negative samples, enhancing the consistency of the representation space. We show the illustration of MD2PR in Figure \ref{fig:md2pr}. 

\subsection{Sentence-level Distillation}\label{sec:sd}

In this section, we discuss the method of knowledge distillation between each sentence pairs. For a query item $q$, there are $n$ documents $d \in D$, including $1$ positive sample and $n-1$ negative samples. A sparse vector $T=(1, 0, 0, \dots, 0)$ of size $n$ is the ground truth labels, indicating whether a query item matches a document. The pseudo-code is shown in Algorithm \ref{algo:sd}.

\begin{algorithm}[t]
	\caption{Sentence-level Distillation\label{algo:sd}}
	\begin{algorithmic}[1]
		\REQUIRE{
			$q$: query item;
			$D$: documents;
			$T$: true labels;
		    $idx$: index of positive samples;}
	    
		\STATE{$S_{ce}$ = ranker(q+d|d $\in$ D);}
		\STATE{$L_{ce}$ = cross\_entropy($S_{ce}$, T);}
		\STATE{$fn_{mask}$ = dynamical\_false\_negative\_filtering($S_{ce}$, idx);}\label{algo:sd1}
		\STATE{$S_{de}$ = retriever(q, d|d $\in$ D);}
		\STATE{$L_{de}$ = cross\_entropy($S_{de}$, T, $fn_{mask}$);}
		\STATE{$L_{sent}$ = KL($S_{ce}$, $S_{de}$);}\label{algo:sd2}
		\STATE{loss = $L_{ce}$ + $L_{de}$ + $L_{sent}$;}
		\STATE{score = $S_{ce}$ + $S_{de}$;}
		\STATE{return loss, score;}
	\end{algorithmic}
\end{algorithm}

For the cross-encoder serving as a ranker, the query item $q$ is concatenated with each document $d \in D$ as input, generating a relevance vector $S_{ce}$ consisting of $n$ scores. Similarly, the dual-encoder retriever takes the query item and document as separate inputs and also generates a relevance vector $S_{de}$. To measure the difference between the true probability distribution and the predicted probability distribution, both the retriever and the ranker utilize cross-entropy as the loss function, shown in Formula \ref{equ:deloss} and \ref{equ:celoss}. These loss functions take the relevance variables and the ground truth labels as inputs to calculate the loss values $L_{ce}$ and $L_{de}$, respectively.

\begin{equation}\label{equ:deloss}
	L_{de}(S_{de}, T) = -\sum^{n}_{1}S_{de}(x_{i})log(T(x_{i}))
\end{equation}

\begin{equation}\label{equ:celoss}
	L_{ce}(S_{ce}, T) = -\sum^{n}_{1}S_{ce}(x_{i})log(T(x_{i}))
\end{equation}

In order to reduce the impact of false negatives, the retriever calculates a false negative mask(line \ref{algo:sd1}) based on the results of the ranker. This mask is obtained through a dynamic false negative filtering method, which will be explained in detail in Section \ref{sec:fn}. Afterwards, we employ the Kullback-Leibler(KL) divergence as the loss function to perform sentence-level distillation(line \ref{algo:sd2}), shown in Formula \ref{equ:sentloss}. 

\begin{equation}\label{equ:sentloss}
	L_{sent}(S_{de},S_{ce}) = \sum S_{de} \cdot log \frac{S_{de}}{S_{ce}}
\end{equation}

The goal is to use the relevance distribution $S_{ce}$ from the ranker to guide the retriever $S_{de}$, enabling it to learn more fine-grained knowledge.

\subsection{Word-level Distillation}\label{sec:word}

To achieve word-level distillation, we consider two important components in a Bert model: the attention and the hidden states. On the one hand, the indirect output of the model is the attention scores, which capture the relationships between each word and other words in the input sequence. On the other hand, the direct output is the hidden states of the last layer of the encoder, which combines contextual semantic information from all layers, with the first position vector being the CLS vector, followed by representations of other words in the sequence. By properly utilizing the hidden states and attention scores, we can achieve effective word-level distillation for retrieval purposes.

As a cross-encoder model serving as the ranker, we choose attention scores as the representation of features. This is because the essence of information retrieval is to determine the relevance between a given query and all the documents. A ranker can directly capture the relevance between words in the query and the document. Attention scores are an important representation of the relationships between words, which are vital for extracting features between query and document. In a Bert model, the hidden state of each layer is computed using the attention mechanism. This design allows it to fully encode input sequences and obtain enough contextual information through the combination of multiple hidden states. Specifically, in the ranker, for each query $q$ of length $|q|$ and each document $d$ of length $|d|$, the concatenation of the query and document is inputted into the model, resulting in an attention matrix with the length of $|q|+|d|$. Each element in the matrix represents the attention score between the corresponding two words in the new sequence.

As a dual-encoder model serving as the retriever, we choose the hidden states to represent the relationships between words in the query or document. This is because the hidden states of last layer in the Bert model contain concrete semantic information. Since output vector of each layer includes both global information and local features, constantly updating in subsequent layers, it eventually generates a hidden state representation for the entire input sequence. In the retriever, given $n$ queries of length $|q|$ and $m$ documents of length $|d|$ for each query, we can concatenate the queries with the documents, resulting in a tensor of shape ($n^2 \times m$, $|q| + |d|$, $|h|$), where $h$ represents the hidden states. By calculating the dot product between vectors, the similarity between every two words in the concatenated sequence is computed. This yields a tensor of shape ($n^2 \times m$, $|q| + |d|$, $|q| + |d|$), representing $n^2 \times m$ new sequences with a length of $|q| + |d|$. As a result, each new sequence corresponds to a matrix of size $(|q| + |d|)^2$, which is used to indicate the attention weights between any two words in the new sequence.

\begin{algorithm}[t]
	\caption{Word-level Distillation\label{algo:wd}}
	\begin{algorithmic}[1]
		\REQUIRE{
			$q$: query item;
			$D$: documents;
			$T$: true labels;
			$idx$: index of positive samples;}
		
		\STATE{$S_{ce}$, $S_{att}$ = ranker(q+d|d $\in$ D);}
		\STATE{$L_{ce}$ = cross\_entropy($S_{ce}$, T);}
		\STATE{$fn_{mask}$ = dynamical\_false\_negative\_filtering($S_{ce}$, idx);}\label{algo:wd1}
		\STATE{$S_{de}$, $S_{sim}$, $p_{mask}$ = retriever(q, d|d $\in$ D);}\label{algo:wd2}
		\STATE{$L_{de}$ = cross\_entropy($S_{de}$, T, $fn_{mask}$);}
		\STATE{$L_{word}$ = MSE($S_{att}$, $S_{sim}$, $p_{mask}$);}\label{algo:wd3}
		\STATE{loss = $L_{ce}$ + $L_{de}$ + $L_{word}$;}
		\STATE{score = $S_{de}$;}
		\STATE{return loss, score;}
	\end{algorithmic}
\end{algorithm}

We show the pseudo-code in Algorithm \ref{algo:wd}. During the forward propagation of the ranker, the attention matrix is used to compute the attention scores $S_{att}$ by taking the mean along the layer and head dimensions. Meanwhile, the retriever takes the input query and documents to generate the corresponding outputs, and processes them to obtain the hidden state matrix $S_{sim}$. Similar to the sentence-level algorithm, we also need to obtain the relevance vectors $S_{ce}$ and $S_{de}$ of the ranker and the retriever, respectively. The cross-entropy loss function is then used to compare the predicted relevance with the ground truth labels $T$.

To achieve word-level distillation, we leverage the fact that the hidden state matrix $S_{sim}$ in the retriever and the attention score matrix $S_{att}$ in the ranker have the same shape. We use both matrices as inputs and optimize them using the mean squared error(MSE) as the loss function(line \ref{algo:wd3}). 

Please note that for the dual-encoder model, its lack of knowledge is mainly focused on the interaction information between query and document. Therefore, we mainly distill the cross-attention scores between query and document from the cross-encoder model, while ignoring the self-attention scores of query or document themselves. As shown in Figure \ref{fig:md2pr}, we extract the information from the upper-right and lower-left parts of the attention scores in the cross-encoder model for distillation. Consequently, we obtain a paired mask matrix $p_{mask}$(line \ref{algo:wd2}) in the retriever to apply to the loss function calculation. The paired mask sets the calculation region to 1 and the other regions to 0, effectively blocking the gradient back propagation with a Softmax function $\varpi$, and thus ignoring the self-attention computation. The specific loss function is shown in Formula \ref{equ:wordloss}.

\begin{equation}\label{equ:wordloss}
	\begin{split}
		L_{word}(S_{att},S_{sim},p_{mask}) = \frac{\sum (S_{att}-\varpi(S_{sim} \cdot p_{mask}))^{2}}{\sum p_{mask}}
	\end{split}
\end{equation}

\subsection{Dynamical False Negative Filtering}\label{sec:fn}

Assessing the relevance between a query and a document lies in a contrastive learning. It maps samples of the same category to similar feature spaces, thereby improving the generalization and robustness. As a result, we can use contrastive learning to compare and differentiate positive and negative instances in the feature space, bringing positive instances closer, while pushing negative instances farther. The two tasks of contrastive learning are loss function designing and positive and negative samples constructing. Therefore, the quality of partitioning positive and negative samples directly affects the training effectiveness. In dense retrieval, finding a highly relevant document sample as a positive one is often done manually, which can guarantee a higher quality. On the other hand, collecting negative samples may be "casual". They can be randomly selected from the entire dataset or within the current batch. Consequently, the selected negative samples may not be actually be true negatives but false negatives, meaning they are highly relevant to the query. This can lead to inaccurate feature representations, resulting in a decrease in model robustness. 

We believe that positive samples should be highly matched to the query, and hence, their scores should be higher than those of other documents. As a result, if there are documents with scores higher than the given positive sample, it indicates a high match with the query and should be considered a positive sample rather a false negative. Therefore, we propose a simple and efficient dynamic method to filter out the false negatives during training. The specific algorithm is shown in Algorithm \ref{algo:dfn}. 

\begin{algorithm}[t]
	\caption{Dynamical False Negative Filtering\label{algo:dfn}}
	\begin{algorithmic}[1]
		\REQUIRE{
			$s$: score achieved by ranker;
			$idx$: index of positive instances;}
		
		\STATE{s = $\varpi$(s);}
		\FOR{i = 1 to n}
		  \IF{i $\neq$ idx and s[i] $\neq$ s[idx]}
		    \STATE{$fn_{mask}$[i] = -$\infty$;}
		  \ELSE
		    \STATE{$fn_{mask}$[i] = 0;}
		  \ENDIF
		\ENDFOR
		\STATE{return $fn_{mask}$;}
	\end{algorithmic}
\end{algorithm}

For a query, if the relevance score $s$ to a document is higher than the relevance score with positive sample, it is considered a false negative and is masked with negative infinity. This mask is applied to its position before participating in distillation, excluding it from calculations. In this way, we can efficiently filter the false negatives with the lowest cost, enhancing the consistency of representation space. Note that the score $s$ should be provided by the ranker since it is a teacher in distillation.

\subsection{MD2PR Analysis}\label{sec:md2pr}

The CLS vector, which represents the sentence-level output, contains the semantic information of the entire sentence. It is suitable for tasks that require comparing the semantics of two sentences; thus, it can better improve the understanding of sentences, and capture their themes and emotions. On the other hand, the word-level hidden states and attention scores, which are the output containing the semantic and interaction information between each word. The features at these two levels have different semantic granularity, and using them together can provide richer information. 

Consequently, we use both sentence-level output and word-level output for distillation. This allows the retriever to learn the knowledge for text matching more comprehensively and deeply, improving its generalization ability and performance on unseen texts. Therefore, we propose the MD2PR, in which the relevant scores, the attention scores, and the hidden states are output during the forward propagation of the ranker and retriever, and then, the distillation of sentence-level and word-level are both utilized to improve the ability of retriever. As a result, the total loss should incorporate both the sentence-level and word-level losses. The loss function is shown in Formula  \ref{equ:totalloss}.

\begin{equation}\label{equ:totalloss}
	\begin{split}
		L_{total} = L_{de} + L_{ce} + L_{sent} + L_{word}
	\end{split}
\end{equation}

It is important to note that when calculating the distillation loss, the ranker should be kept fixed for stable training. Furthermore, whether it is sentence-level distillation, word-level distillation, or the combination of both levels, the cross-encoder ranker is only used during the training process and will not be used in the subsequent encoding and retrieval. This is because the role of the ranker is to obtain more comprehensive semantic information through attention calculations on a large amount of training data and teach this interaction information to the retriever. The subsequent encoding and retrieval belong to the inference stage, during which if the ranker is used, we have calculate the attention scores between the given query and all documents in the dataset, which is computationally expensive and not feasible for real-time response. On the other hand, for the retriever that has already learned the knowledge from the ranker, it can pre-encode all the documents. Given a query, only the encoding of the query needs to be calculated and then used to compute the similarity with the document encodings. This significantly reduces the computational cost and latency from input to output, making it particularly suitable for real-time applications like search engines.

\section{Experiments}\label{sec:expr}

\subsection{Experimental Preparation}

\subsubsection{Metrics}

Information retrieval measures the degree of match between a query and a document, where document with higher matching scores is ranked higher, resulting in a better performance. The two commonly used metrics are Mean Reciprocal Rank($MRR$) and $Recall$.

$MRR$ calculates the average reciprocal rank when retrieving the first relevant document. The goal is to achieve the most relevant item and have it ranked higher. The calculation of $MRR$ is shown in Formula \ref{equ:mrr}. 

\begin{equation}\label{equ:mrr}
	MRR = \frac{1}{|Q|}\sum_{i=1}^{|Q|}\frac{1}{rank_{i}}
\end{equation}

$Recall$ measures the query recall rate of the retriever. It is defined as the proportion of the relevent results retrieved to all the relevent documents. The calculation of $Recall$ is shown in Formula \ref{equ:recall}. 

\begin{equation}\label{equ:recall}
	Recall = \frac{TP}{TP+FN}
\end{equation}

In the formula, $TP$ represents true positives, which are the relevant results that can be retrieved. $FP$ represents false negatives. $TP$ + $FN$ represents all existing relevant results.

It is important to note that $MRR$ and $Recall$ are interrelated metrics in information retrieval. Generally, it is desirable to improve $MRR$ while ensuring a high $Recall$. In our experiments, we use $MRR@k$ and $R@k$ as evaluation metrics, where $k$ represents the number of top-ranked matching documents to consider for each query item when calculating the corresponding metrics.

\subsubsection{Datasets}

We selected two classic datasets commonly used in the field of text retrieval and question answering: MS-MARCO and Natural Questions. MS-MARCO\cite{nguyen2016ms} is a dataset initially introduced by Microsoft in 2016 for reading comprehension, and later became widely used in the information retrieval domain. This dataset consists of 8.8 million paragraphs extracted from web pages collected by Bing, corresponding to 1 million true query items. Each query item is associated with a passage labeled as relevant, serving as positive sample. The training and testing sets are divided into 5,023,939 and 6,837 query items, respectively.

Natural Questions\cite{kwiatkowski2019natural} was initially designed for open-domain question answering. The questions in this dataset are extracted from user queries in Google Search, and the answers are manually identified passages from Wikipedia articles. The training and testing sets are comprised of about 58,812 and 3,610 samples, respectively.

The information statistics of these two datasets are shown in Table \ref{tab:dataset}.

\begin{table}
	\caption{Characteristics of Datasets}
	\label{tab:dataset}
	\begin{tabular}{cccccccc}
		\toprule
		Name & Training & Testing & Doc & Query & Doc\\
		 & Num & Num & Num & Len & Len\\
		\midrule
		MS-MARCO & 5023939 & 6837 & 8841823 & 5.97 & 56.58\\
		Nat. Quest. & 58812 & 3610 & 21015324 & 9.20 & 100.0\\
		\bottomrule
	\end{tabular}
\end{table}

\subsubsection{Baselines}

The experiment was implemented using Python 3.8 and ran on Ubuntu 20.04 with a 2.50GHz CPU and an Nvidia RTX 3090 GPU. The optimizer  was Adam Weight Decay Regularization(Adamw). The learning rate for the retriever ranged from 5e-6 to 10e-6, with a warm-up ratio of 0.1. 

To evaluate our model, we selected 11 baseline models, including 2 traditional sparse retrieval models BM25 and doc2query, 2 sparse retrieval models enhanced with deep language models DeepCT and DocT5Query, and 7 dense retrieval models, DPR, COIL, ANCE, RocketQA, RocketQA v2, Condenser, and CoCondenser.

\subsection{Experimental Results}

\subsubsection{Evaluation with Baselines}

In this study, our model MD2PR utilized the Bert, which was also used in COIL, as the retriever and the DeBERTa\cite{he2020deberta} as the ranker. The models were trained and validated on both datasets with hyper-parameter tuning, and the results are summarized in Tables \ref{tab:mm} and \ref{tab:nq}. 

\begin{table*}
	\caption{Results over MS-MARCO and Natural Questions(\%, - represents the experimental results are unavailable)}
	\label{tab:mm}
	\begin{tabular}{ccccccc}
		\toprule
		\multirow{2}{*}{Model} & \multicolumn{3}{c}{MS-MARCO} & \multicolumn{3}{c}{Natural Questions} \\
		& $MRR@10$ & $R@50$ & $R@1000$ & $R@5$ & $R@20$ & $R@100$ \\
		\midrule
		BM25\cite{robertson2009probabilistic} & 18.7 & 59.2 & 85.7 & 18.7 & 59.2 & 85.7 \\
		doc2query\cite{nogueira2019document} & 21.5 & 64.4 & 89.1 & - & - & -\\
		DocT5Query\cite{nogueira2019doc2query} & 27.7 & 75.6 & 94.7 & - & - & - \\
		DeepCT\cite{dai2019context} & 24.3 & 69.0 & 91.0 & - & - & - \\
		DPR\cite{karpukhin2020dense} & 29.9 & - & 92.8 & - & 78.4 & 85.3 \\
		ANCE\cite{xiong2020approximate} & 33.0 & - & 95.9 & - & 81.9 & 87.5 \\
		COIL\cite{gao2021coil} & 35.5 & - & 96.3 & - & - & - \\
		RocketQA\cite{qu2020rocketqa} & 37.0 & 85.5 & 97.9 & 74.0 & 82.7 & 88.5\\
		Condenser\cite{gao2021condenser} & 36.6 & - & 97.4 & - & 83.2 & 88.4\\
		RocketQA v2\cite{ren2021rocketqav2} & \textbf{38.2} & \textbf{86.2} & 98.1 & 75.1 & 83.7 & 89.0\\
		CoCondenser\cite{gao2021unsupervised} & 38.2 & - & \textbf{98.4} & 75.8 & 84.3 & 89.0\\
		AR2\cite{zhang2021adversarial} & 39.5 & 87.8 & \textbf{98.6} & 77.9 & 86.0 & 90.1\\
		MD2PR(full) & 36.9 & 83.1 & 97.4 & \textbf{75.9} & \textbf{84.8} & \textbf{89.6} \\
		\bottomrule
	\end{tabular}
\end{table*}

MD2PR(full) represents the complete version of the MD2PR model, which includes both the sentence-level and word-level distillation as well as the dynamic negative sample filtering method. The baseline model metrics are obtained from the experimental results published in the existing papers. For the data of Natural Questions, the experiments primarily focused on dense retrieval models, so the models used for comparison are also dense retrieval models.

As can be seen, similar to other dense retrieval models, our MD2PR model proposed in this paper significantly outperforms traditional sparse retrieval models such as BM25 and doc2query, as well as enhanced sparse retrieval models like DocT5Query and DeepCT. Furthermore, compared to the dense retrieval models, MD2PR also demonstrates good performance. On the MS-MARCO dataset, the MD2PR model achieved a 1.4\% higher $MRR@10$ metric compared to the state-of-the-art COIL model, and a 1.1\% higher $R@1000$ metric than COIL. On the Natural Questions dataset, MD2PR achieved a 0.8\% higher $R@5$ metric than the dense retrieval model RocketQA v2, a 0.5\% higher $R@20$ metric than the dense retrieval model CoCondenser, and a 0.6\% higher $R@100$ metric than CoCondenser. 

The experiments also indicates that when $k$ is small, the difficulty of retrieving positive samples increases as fewer documents are screened out. This results in a greater differentiation in performance among different models, reflecting their varying capabilities. However, as $k$ increases and more documents are considered as candidates, the probability of positive samples appearing in the top-$k$ also increases. In this case, the performance differences between models decrease.

\subsubsection{Ablation Studies}

To further investigate the effectiveness of the 3 proposed methods in this paper and their interactions, we conducted ablation experiments using COIL as the baseline model. We added or removed different methods and evaluated them on the MS-MARCO dataset using the $R@1000$ metric. The results of these experiments are presented in Table \ref{tab:ablation}.

\begin{table}
	\caption{Ablation Studies over MS-MARCO(\%, sd: sentence-level distillation; wd: word-level distillation; fnf: dynamical false negative filtering)}
	\label{tab:ablation}
	\begin{tabular}{cccc}
		\toprule
		Method & $MRR@10$ & $R@50$ & $R@1000$\\
		\midrule
		MD2PR(basic) & 35.5 & 81.2 & 96.3\\
		MD2PR(basic+sd) & 36.1 & 81.5 & 96.5\\
		MD2PR(basic+wd) & 36.5 & 82.6 & 96.9\\
		MD2PR(basic+fnf) & 36.2 & 81.7 & 96.6\\
		MD2PR(basic+sd+wd) & 36.7 & 82.9 & 97.0\\
		MD2PR(basic+sd+wd+fnf) & 36.9 & 83.1 & 97.4\\
		\bottomrule
	\end{tabular}
\end{table}

From the experimental results, it can be observed that introducing the methods proposed in this paper leads to effective improvements compared to the basic model. Actually, the basic model utilized a Bert-based dual-encoder model, and using a larger model for knowledge distillation can easily enhanced its performance. Among the proposed methods, the word-level distillation demonstrates a more significant improvement. This is because word-level distillation provides a finer-grained distillation process, not only transferring relationships between internal words to the retrieval but also capturing the relationships between query and document. The utilization of dynamic negative sample filtering method also outperforms sentence-level distillation, indicating that the problem of false negatives has been addressed, while the issue of sentence-level distillation might not fully exploit the performance due to the presence of false negatives.

The combined use of sentence-level and word-level distillation further improved the model performance compared to using them individually. This is because the sentence-level and word-level distillation can capture the representation of the relationship between query and document in different dimensions, achieving a mutually reinforcing effect. Moreover, by incorporating dynamic negative sample filtering method on the distillation at both levels, inappropriate data is filtered out, making the respective feature spaces more complete and effective. As a result, the interaction between the two levels of distillation can reflect the inherent characteristics of the data more accurately, leading to a better retrieval performance overall.

\subsubsection{Different Models of Retriever and ranker}

Different versions of Bert models, such as  ALBert\cite{lan2019albert}, DeBERTa, and ERNIE\cite{zhang2019ernie,sun2020ernie,sun2021ernie}, having different size, can be used on both the ranker and retriever. Typically, the size of the ranker is not smaller than that of the retriever to ensure the effectiveness of distillation. The experimental results, as shown in Table \ref{tab:diffmodel}, were obtained by using various scales of these models.

\begin{table*}
	\caption{Different Models of Retriever and ranker}
	\label{tab:diffmodel}
	\begin{tabular}{cccccccc}
		\toprule
		Retriever & Retriever & Ranker & Ranker & $R@1000$(\%) & $R@100$(\%) & Training Time(h) & Latency(ms)\\
		 & size(billion) & & size(billion) & on MS-MARCO & on N.Q. &  & on MS-MARCO \\
		\midrule
		Albert(base)-v2 & 0.012 & Bert(base) & 0.109 & 89.8 & 80.1 & 16 & 7\\
		Albert(base)-v2 & 0.012 & DeBERTa(xlarge)-v2 & 0.885 & 97.2 & 89.3 &  461 & 7\\
		Albert(base)-v2 & 0.012 & DeBERTa(large)-v3 & 0.434 & 97.5 & 89.7 & 210 & 7\\
		Albert(large)-v2 & 0.018 & Bert(base) & 0.109  & 90.5 & 80.9 & 29 & 9\\
		Albert(xlarge)-v2 & 0.059 & Bert(base) & 0.109 & 91.7 & 81.6 & 53 & 16\\
		Bert(base) & 0.109 & ERNIE(base) & 0.109 & 90.7 & 81 & 31 & 31\\
		Bert(base) & 0.109 & ERNIE(large) & 0.335 & 93 & 83.6 & 73 & 31\\
		Bert(base) & 0.109 & Bert(large) & 0.335 & 91.8 & 81.8 & 54 & 31\\
		Bert(base) & 0.109 & DeBERTa(base) & 0.139 & 92.3 & 82.5 & 63 & 31\\
		Bert(base) & 0.109 & DeBERTa(large) & 0.405 & \textbf{97.4} &  \textbf{89.6} & \textbf{134} & \textbf{31} \\
		Bert(base) & 0.109 & DeBERTa(large)-v3 & 0.434 & 97.3 & 89.6 & 234 & 31\\
		Bert(base) & 0.109 & Albert(xxlarge)-v2 & 0.223 & 95.1 & 86.2 & 120 & 31\\
		Bert(large) & 0.335 & DeBERTa(large) & 0.405 & \textit{97.5} & \textit{89.8} & \textit{510} & \textit{63}\\
		\bottomrule
	\end{tabular}
\end{table*}

We can see that all the distillations over different models are effective. Generally, a larger ranker may result in a better distillation performance. As an example, when the retriever used the Albert(base) with 0.012 billion parameters, using DebertTa(xlarge) with 0.8 billion parameters rather Bert(base) with 0.1 billion parameters can raise the $R@1000$ on MS-MARCO from 89.8\% to 97.2\%, and can raise the $R@100$ on Natural Questions from 80.1\% to 89.3\%. But on the other hand, if the size difference between the models used in the retriever and ranker is not significant, it leads to better model performance. For example, when the retriever was Bert(base) and the ranker is DeBERTa(large), the $R@1000$ metric on the MS-MARCO dataset is 97.4\%, with a parameter difference of 3.7 times between the two models. However, if the retriever is AlBert(base) and the ranker is DeBERTa(xlarge), the parameter difference is 75.7 times, but the $R@1000$ is only 97.2\%. Clearly, the difference in model size between the ranker and retriever does not necessarily result in better model performance. Intuitively, using models with a significantly difference for distillation may lead to information loss, as the smaller model might not be able to capture the complex features and information present in the larger model. Additionally, models with a significantly difference might have different pre-training objectives and training processes, which could adversely affect the effectiveness of the distillation process.

We also shown the training time of 3 epochs under different model combinations in the Column 7 of Table \ref{tab:diffmodel}. From the experimental results, it can be obviously observed that as the model parameters of the retriever and ranker gradually increased, the training time also increased while the model performance improved significantly. However, when both the retriever and ranker used large models, shown in the last row, the training time is nearly 4 times that of the baseline model, bold in the 10th row, but the improvement in $R@1000$ is only 0.1\%. This indicates that increasing the model size does not always lead to continuous performance improvement. Under the current parameter settings and dataset, the model may encounter overfitting issues.

Moreover, we also recorded the search latency on MS-MARCO in Column 8. Generally, since we only used the retriever in this process, the latency mostly depended on the size of the retriever. This was validated with our experiment shown in column 8, which is the average search latency of random selected 100 queries from MS-MARCO. As can be seen, the latency increased following the size of retriever, regardless of any changes in the ranker.

\section{Conclusions}\label{sec:conclude}

To address the problems of high computation cost for cross-encoder and insufficient interaction capability of dual-encoder in dense retrieval, we proposed a model named MD2PR based on sentence-level and word-level knowledge distillation. In this model, sentence-level distillation was performed using the vector corresponding to the CLS token, while word-level distillation was conducted by align attention scores and hidden states. Additionally, a dynamic filtering method of false negatives was introduced to obtain better representation space. The effectiveness of these 3 methods were explored and validated by conducting experiments on 2 classic information retrieval datasets, MS-MARCO and Natural Questions, using standard evaluation metrics, the $MRR$ and the $Recall$. The results are evaluated against 11 baseline models, including sparse and dense retrieval models. The experimental results demonstrate that MD2PR outperformed the sparse retrieval model significantly, and outperformed the state-of-the-art dense retrieval model COIL with a 1.4\% improvement in $MRR@10$ and a 1.1\% improvement in $R@1000$ on the MS-MARCO dataset. On the Natural Questions dataset, MD2PR achieved an 0.8\% higher $R@5$ than the RocketQA v2 model and a 0.6\% higher $R@100$ than the CoCondenser model. Furthermore, since we only used the dual-encoder model in the inference retrieval, the computational cost was significantly reduced. 

\section*{Acknowledgments}
This should be a simple paragraph before the References to thank those individuals and institutions who have supported your work on this article.



%\begin{thebibliography}{1}
\bibliographystyle{ACM-Reference-Format}

\bibliography{sample-base}

%\bibliography{cas-refs}

%\bibitem{ref1}
%{\it{Mathematics Into Type}}. American Mathematical Society. [Online]. Available: https://www.ams.org/arc/styleguide/mit-2.pdf
%
%\bibitem{ref2}
%T. W. Chaundy, P. R. Barrett and C. Batey, {\it{The Printing of Mathematics}}. London, U.K., Oxford Univ. Press, 1954.
%
%\bibitem{ref3}
%F. Mittelbach and M. Goossens, {\it{The \LaTeX Companion}}, 2nd ed. Boston, MA, USA: Pearson, 2004.
%
%\bibitem{ref4}
%G. Gr\"atzer, {\it{More Math Into LaTeX}}, New York, NY, USA: Springer, 2007.
%
%\bibitem{ref5}M. Letourneau and J. W. Sharp, {\it{AMS-StyleGuide-online.pdf,}} American Mathematical Society, Providence, RI, USA, [Online]. Available: http://www.ams.org/arc/styleguide/index.html
%
%\bibitem{ref6}
%H. Sira-Ramirez, ``On the sliding mode control of nonlinear systems,'' \textit{Syst. Control Lett.}, vol. 19, pp. 303--312, 1992.
%
%\bibitem{ref7}
%A. Levant, ``Exact differentiation of signals with unbounded higher derivatives,''  in \textit{Proc. 45th IEEE Conf. Decis.
%Control}, San Diego, CA, USA, 2006, pp. 5585--5590. DOI: 10.1109/CDC.2006.377165.
%
%\bibitem{ref8}
%M. Fliess, C. Join, and H. Sira-Ramirez, ``Non-linear estimation is easy,'' \textit{Int. J. Model., Ident. Control}, vol. 4, no. 1, pp. 12--27, 2008.
%
%\bibitem{ref9}
%R. Ortega, A. Astolfi, G. Bastin, and H. Rodriguez, ``Stabilization of food-chain systems using a port-controlled Hamiltonian description,'' in \textit{Proc. Amer. Control Conf.}, Chicago, IL, USA,
%2000, pp. 2245--2249.

%\end{thebibliography}

\vfill

\end{document}